\documentclass[aps,prd,nofootinbib,showpacs,superscriptaddress,twocolumn,10pt]{revtex4}
\usepackage{graphicx}
\usepackage{dcolumn}
\usepackage{bm}
\usepackage{amssymb}
\usepackage{latexsym}
\usepackage{booktabs}
\usepackage[colorlinks, linkcolor=blue, citecolor=blue, urlcolor=blue]{hyperref}

\newcommand{\be}{\begin{equation}}
\newcommand{\ee}{\end{equation}}
\newcommand{\bq}{\begin{eqnarray}}
\newcommand{\eq}{\end{eqnarray}}

\begin{document}

\title{Interacting model of new agegraphic dark energy: observational constraints and age problem}

\author{Yun-He Li}
\affiliation{Department of Physics, College of Sciences,
Northeastern University, Shenyang 110004, China}
\author{Jing-Zhe Ma}
\affiliation{Department of Physics, College of Sciences,
Northeastern University, Shenyang 110004, China}
\author{Jing-Lei Cui}
\affiliation{Department of Physics, College of Sciences,
Northeastern University, Shenyang 110004, China}
\author{Zhuo Wang}
\affiliation{Department of Physics, College of Sciences,
Northeastern University, Shenyang 110004, China}
\author{Xin Zhang\footnote{Corresponding author}}
\email{zhangxin@mail.neu.edu.cn} \affiliation{Department of Physics,
College of Sciences, Northeastern University, Shenyang 110004,
China} \affiliation{Center for High Energy Physics, Peking
University, Beijing 100080, China}

\begin{abstract}
Many dark energy models fail to pass the cosmic age test because of
the old quasar APM 08279+5255 at redshift $z=3.91$, the $\Lambda$CDM
model and holographic dark energy models being no exception. In this
paper, we focus on the topic of age problem in the new agegraphic
dark energy (NADE) model. We determine the age of the universe in
the NADE model by fitting the observational data, including type Ia
supernovae (SNIa), baryon acoustic oscillations (BAO) and the cosmic
microwave background (CMB). We find that the NADE model also faces
the challenge of the age problem caused by the old quasar APM
08279+5255. In order to overcome such a difficulty, we consider the
possible interaction between dark energy and dark matter. We show
that this quasar can be successfully accommodated in the interacting
new agegraphic dark energy (INADE) model at the $2\sigma$ level
under the current observational constraints.
\end{abstract}

\pacs{95.36.+x, 98.80.Es, 98.80.-k} \maketitle


\section{Introduction}\label{sec1}

At its present stage, our universe is undergoing an accelerated
expansion, which has been confirmed by many astronomical
observations, such as, type Ia supernovae (SNIa) \cite{11Riess98},
large scale structure (LSS) \cite{12Tegmark04} and the cosmic
microwave background (CMB) \cite{13Spergel03}, among others. All
these observations indicate the existence of a ``dark energy'' with
negative pressure. The most important theoretical candidate for dark
energy is the cosmological constant, $\Lambda$, which fits the
observations well, but is plagued with some severe theoretical
difficulties, such as the so called ``fine-tuning'' and ``cosmic
coincidence'' problems \cite{14dereview}. During the past decade, in
order to unveil its nature, theorists have proposed many
phenomenological models of dark energy, e.g., quintessence
\cite{15quintessence}, $k$-essence \cite{16kessence}, tachyon
\cite{19tachyon}, phantom \cite{18phantom}, quintom
\cite{21quintom}, braneworld \cite{brane}, and Chaplygin gas
\cite{cg}. Further, the possibility that dark energy might interact
with dark matter, owing to their unknown nature, has also been
seriously considered in many works to help solve the cosmic
coincidence problem \cite{intde1} and the cosmic doomsday problem
\cite{intde2}. For reviews of dark energy, see, e.g.,
\cite{Sahni:1999gb}.

In recent years, it has been found that many dark energy models get
into trouble when tested by some old high redshift objects (OHROs).
It is obvious that the universe cannot be younger than its
constituents, so the age of some astronomical objects (at some
redshift), if measured accurately, can be used to test cosmological
models according to this simple age principle. There have been some
OHROs discovered, including, the $3.5$ Gyr old galaxy LBDS 53W091 at
redshift $z=1.55$~\cite{27Dunlop:1996mp} and the $4.0$ Gyr old
galaxy LBDS 53W069 at redshift $z=1.43$~\cite{28Dunlop:1998tm}. In
particular, the old quasar APM $08279+5255$ at redshift $z=3.91$ is
an important one, which has been used as a ``cosmic clock'' to
constrain cosmological models. Its age is estimated to be $2.0-3.0$
Gyr \cite{29Hasinger:2002wg}. These three OHROs at $z=1.43$, 1.55
and 3.91 have been used to test many dark energy models, including,
the $\Lambda$CDM model~\cite{31Alcaniz:1999kr}, the general EoS dark
energy model~\cite{32Dantas:2006dy}, the scalar-tensor quintessence
model~\cite{33Capozziello:2007gr}, the $f(R)=\sqrt{R^2-R_0^2}$
model~\cite{34Movahed:2007cs}, the DGP braneworld
model~\cite{35Movahed:2007ie}, the power-law parameterized
quintessence model~\cite{36Rahvar:2006tm}, the Yang-Mills condensate
model~\cite{37Tong:2009mu}, the holographic dark energy
model~\cite{Wei:2007ig}, and the agegraphic dark energy
model~\cite{38Zhang:2007ps}. These investigations show that the two
OHROs at $z=1.43$ and 1.55 can be easily accommodated in most dark
energy models, whereas the OHRO at $z=3.91$ cannot, even in the
$\Lambda$CDM model~\cite{31Alcaniz:1999kr} and the holographic dark
energy model ~\cite{Wei:2007ig}. In this paper, we will investigate
the cosmic age problem in the new agegraphic dark energy (NADE)
model. We will show that the NADE model also faces the challenge of
such an age problem. In order to escape from the cosmic age crisis,
we consider the possible interaction between dark energy and dark
matter. We will check whether the age problem can be solved
successfully in the interacting new agegraphic dark energy (INADE)
model. Of course, our discussions are based on the current
observational constraints on the models. So, we will first place
observational constraints on the NADE and INADE models, and then
discuss the cosmic age problem.



\section{The new agegraphic dark energy model with interaction}\label{sec2}

In this section, we describe the INADE model in a flat universe.
Many theorists believe that we cannot entirely understand the nature
of dark energy before a complete theory of quantum gravity is
established \cite{39Witten:2000zk}. In the circumstance that a full
theory of quantum gravity is not yet available, it is more realistic
to consider the possible cosmological consequences of some
fundamental principles of quantum gravity. The holographic principle
\cite{holop} is commonly believed to be such a principle, so it is
expected to play an essential role in investigating dark energy
\cite{Cohen99}. Along this line, a model of holographic dark energy
has been proposed \cite{41Li:2004rb} (see also, e.g.,
\cite{holofitzx} and \cite{holoext}). The agegraphic dark energy
model \cite{42Cai:2007us} is constructed in light of the
K\'{a}rolyh\'{a}zy relation \cite{43uncertainty} and corresponding
energy fluctuations of space-time. Actually, it has been proven that
the agegraphic dark energy scenario is also a kind of holographic
model of dark energy \cite{42Cai:2007us}. In such a holographic
model, the UV problem of dark energy is converted into an IR
problem, since the dark energy density is inversely proportional to
the square of the IR length scale, $\rho_{de}\sim L^{-2}$. In the
old version of the agegraphic dark energy model \cite{42Cai:2007us},
the IR cutoff is chosen as the age of the universe, $t$ (here it
should be pointed out that the speed of light has already been taken
to be 1, so time and length have the same dimension). However, there
are some inner inconsistencies in this model; for details see
\cite{47Wei:2007ty}. Therefore, in this paper, we only discuss the
new version of the agegraphic dark energy model \cite{47Wei:2007ty}.

In the NADE model, the IR cutoff is chosen to be the conformal age
of the universe,
 \be\label{eq1}
   \eta\equiv\int_0^t\frac{d\tilde{t}}{a}=\int_0^a\frac{d\tilde{a}}{H\tilde{a}^2},
 \ee
so the energy density of NADE is
 \be\label{eq2}
   \rho_q=\frac{3n^2m_p^2}{\eta^2},
 \ee
where $n$ is a numerical parameter and $m_p$ is the reduced Planck
mass.

If we consider a spatially flat FRW universe containing agegraphic
dark energy and dark matter, the corresponding Friedmann equation is
 \be\label{eq3}
   H^2=\frac{1}{3m_p^2}\left(\rho_m+\rho_q\right),
 \ee
or equivalently,
 \be\label{eq4}
   E(z)\equiv {H(z)\over H_0}=\left(\Omega_{m0}(1+z)^3\over1-\Omega_q(z)\right)^{1/2},
 \ee
where $\Omega_{m0}$ is the present fractional dark matter density
and $\Omega_q\equiv\rho_q/(3m_p^2H^2)$. Note that, for simplicity,
in this paper we neglect the contributions from baryons and
radiation. From (\ref{eq2}), it is easy to find that
 \be\label{eq5}
   \Omega_q=\frac{n^2}{H^2\eta^2}.
 \ee
Obviously, $\Omega_m\equiv\rho_m/(3m_p^2H^2)=1-\Omega_q$ from
(\ref{eq3}). By using (\ref{eq1}), (\ref{eq2}), (\ref{eq3}) and
(\ref{eq5}) and the energy conservation equation
$\dot{\rho}_m+3H\rho_m=0$, we obtain the equation of motion for
$\Omega_q$:
 \be\label{eq6}
   \Omega_q^\prime=\Omega_q\left(1-\Omega_q\right)\left(3-\frac{2}{na}\sqrt{\Omega_q}\right),
 \ee
where the prime denotes the derivative with respect to $x\equiv\ln
a$. Since $\frac{d}{dx}=-\left(1+z\right)\frac{d}{dz}$, we get
 \be\label{eq7}
   \frac{d\Omega_q}{dz}=-\frac{\Omega_q}{1+z}\left(1-\Omega_q\right)\left(3-\frac{2\left(1+z\right)}{n}\sqrt{\Omega_q}\right).
 \ee
From the energy conservation equation
$\dot{\rho}_q+3H(\rho_q+p_q)=0$, as well as (\ref{eq2}) and
(\ref{eq5}), it is easy to find that the equation-of-state (EoS)
parameter of the NADE model is given by
 \be\label{eq8}
   w_q=-1+\frac{2}{3na}\sqrt{\Omega_q}.
 \ee
The NADE model has been studied extensively; see, e.g.,
\cite{48Wei:2007xu,49nadeext}. In the following, we shall extend the
NADE model by considering the interaction between dark energy and
dark matter.

Assuming that dark energy and dark matter exchange energy through
the interaction term $Q$, the continuity equations become
 \be\label{eq9}
   \dot{\rho}_q+3H\left(\rho_q+p_q\right)=-Q
 \ee
 and
 \be\label{eq10}
   \dot{\rho}_m+3H\rho_m=Q.
 \ee
Owing to the lack of any knowledge of the micro-origin of the
interaction, we simply follow other work on interacting dark energy
and parameterize the interaction term generally as
$Q=3H(\alpha\rho_q+\beta\rho_m)$, where $\alpha$ and $\beta$ are
dimensionless coupling constants. In order to reduce both complexity
and the number of parameters, one often considers the following
three cases: (i) $\alpha=b$ and $\beta=0$, denoted as INADE1, (ii)
$\alpha=0$ and $\beta=b$, denoted as INADE2, and (iii)
$\alpha=\beta=b$, denoted as INADE3. Note that according to our
convention $b > 0$ means that dark energy decays to dark matter,
while $b < 0$ means that dark matter decays into dark energy. In
cases (i) and (iii), a negative value of $b$ would lead to the
unphysical consequence that $\rho_m$ becomes negative in the distant
future. For negative $b$ values in case (ii), no such difficulty
exists. In \cite{50Pavon:2007gt} it is argued from the
thermodynamical view that the second law of thermodynamics strongly
favors that dark energy decays into dark matter. So, in general, $b$
is taken to be positive.

However, recently, it has been found that observations may favor the
case of dark matter decaying into dark energy
\cite{50Pereira:2008at,50Costa:2009mv}. In particular, in
\cite{50Cai:2009ht}, in a way independent of specific interacting
forms, the authors fitted for the interaction term $Q$ using
observations. They found that $Q$ is likely to cross the
non-interacting line $(Q=0)$, i.e., the sign of the interaction $Q$
is changed, around $z=0.5$. This raises a remarkable challenge to
the interacting models, since the general phenomenological forms of
interaction, as shown in the above, do not provide the possibility
of changing signs. As noted in \cite{50Cai:2009ht}, more general
forms of interaction should be considered. For this reason, a new
form of interaction, (iv) $\alpha=-\beta=b$, denoted as INADE4, was
considered in \cite{51Sun:2010vz}. Obviously, for this case, in the
early stage, since $\rho_m>\rho_q$, $Q$ is negative. However, $Q$
may change from negative to positive when the expansion of the
universe changes from decelerating to accelerating. The parameter
$b$ in this case is also assumed to be positive, since negative $b$
would lead to a negative $\rho_m$ in the distant future.

For clarity, we denote the aforementioned interation term $Q$ as
 \begin{eqnarray}\label{eq11}
   Q = \left\{\begin{array}{ll}
     3bH\rho_q,         \\
     3bH\rho_m,         \\
     3bH(\rho_q+\rho_m),\\
     3bH(\rho_q-\rho_m).
   \end{array}\right.
 \end{eqnarray}
Note that in cases (i), (iii) and (iv) the parameter $b$ is always
assumed to be positive in the literature. However, in the present
work, instead of making such an assumption, we permit this variable
to be completely free and allow the observational data to tell us
the true story, no matter if the ultimate fate of the universe is
ridiculous or not.

Differentiating (\ref{eq5}) with respect to $\ln a$ and using
(\ref{eq1}), we get
 \be\label{eq12}
   \Omega_q^\prime=\Omega_q\left(-2\frac{\dot{H}}{H^2}-\frac{2}{na}\sqrt{\Omega_q}\right).
 \ee
Differentiating (\ref{eq3}) with respect to time $t$ and combining
(\ref{eq1}), (\ref{eq5}), (\ref{eq9}) and (\ref{eq10}), we can
easily find that
 \be\label{eq13}
   -\frac{\dot{H}}{H^2}=\frac{3}{2}\left(1-\Omega_q\right)+\frac{\Omega_q^{3/2}}{na}-\frac{Q}{6m_p^2 H^3}.
 \ee
Therefore, we obtain the equation of motion for $\Omega_q$,
 \be\label{eq14}
   \Omega_q^\prime=\Omega_q\left[\left(1-\Omega_q\right)\left(3-\frac{2}{na}\sqrt{\Omega_q}\right)-Q_1\right],
 \ee
or equivalently,
 \be\label{eq15}
   \frac{d\Omega_q}{dz}=-\frac{\Omega_q}{1+z}\left[\left(1-\Omega_q\right)\left(3-\frac{2\left(1+z\right)}{n}\sqrt{\Omega_q}\right)-Q_1\right],
 \ee
where
 \be\label{eq16}
   Q_1\equiv\frac{Q}{3m_p^2 H^3}.
 \ee
From (\ref{eq5}) and (\ref{eq9}), we get the EoS parameter of dark
energy:
 \be\label{eq17}
   w_q=-1+\frac{2}{3na}\sqrt{\Omega_q}-Q_2,
 \ee
where
 \be\label{eq18}
   Q_2\equiv\frac{Q}{3H\rho_q}.
 \ee
It is convenient to define the effective EoS parameters for dark
energy and dark matter as
 \be\label{eq19}
   {w_q^{\left(e\right)}}=w_q+\frac{Q}{3H\rho_q}
 \ee
 and
 \be\label{eq20}
   {w_m^{\left(e\right)}}=-\frac{Q}{3H\rho_m}.
 \ee
According to the definition of the effective EoS parameters, the
continuity equations for dark energy and dark matter can be
re-expressed in the form of energy conservation:
 \be\label{eq21}
   \dot{\rho}_q+3H(1+{w_q^{\left(e\right)}})\rho_q=0
 \ee
 and
 \be\label{eq22}
   \dot{\rho}_m+3H(1+{w_m^{\left(e\right)}})\rho_m=0.
 \ee
Taking the aforementioned four cases of interaction, one can obtain
 \begin{eqnarray}\label{eq23}
   {w_m^{\left(e\right)}} =\left\{\begin{array}{ll}
     -b\frac{\Omega_q}{1-\Omega_q}                ~~~ &~~~Q=3bH\rho_q,           \\\\
     -b                                           ~~~ &~~~Q=3bH\rho_m,           \\\\
     -b\left(1+\frac{\Omega_q}{1-\Omega_q}\right) ~~~ &~~~Q=3bH(\rho_q+\rho_m),  \\\\
     -b\left(-1+\frac{\Omega_q}{1-\Omega_q}\right)~~~ &~~~Q=3bH(\rho_q-\rho_m).
   \end{array}\right.
 \end{eqnarray}
Now, the Friedmann equation can be expressed as
 \be\label{eq24}
   H(z)=H_0E(z),
 \ee
where
 \be\label{eq25}
   E(z)=\left[\frac{(1-\Omega_{q0})e^{3\int_0^z{1+w_m^{(e)}\over 1+\tilde{z}}d\tilde{z}}}{1-\Omega_q(z)}\right]^{1/2}.
 \ee
In the above equation, $\Omega_q(z)$ can be obtained by numerically
solving (\ref{eq15}) with the initial condition
$\Omega_q(z_{ini})=n^2(1+z_{ini})^{-2}/4$ at $z_{ini}=2000$
\cite{48Wei:2007xu}. While this initial condition is obtained from a
NADE model without interaction in the matter-dominated epoch, it is
still suitable to an interacting model of NADE because the
contribution of dark energy to the cosmological evolution is
negligible in the matter-dominated epoch and therefore its impact
can be ignored during that time; for details
see~\cite{Zhang:2009qa}.


\section{Observational constraints}\label{sec3}

In this section, we place observational constraints on the NADE and
INADE models. For the data, we will use the combination of SNIa, CMB
and BAO.

First, we consider the $557$ Union2 SNIa compiled in
\cite{52Amanullah:2010vv}. The theoretical distance modulus is
defined as
 \be\label{eq26}
   \mu_{th}(z_i)\equiv5\log_{10} D_L(z_i)+\mu_0,
 \ee
where $\mu_0\equiv42.38-5\log_{10} h$ with $h$ the Hubble constant
$H_0$ in units of 100 km s$^{-1}$ Mpc$^{-1}$, and the Hubble-free
luminosity distance
 \be\label{eq27}
   D_L(z)=(1+z)\int_0^z \frac{d\tilde{z}}{E(\tilde{z};{\bf p})},
 \ee
where $E\equiv H/H_0$, and ${\bf p}$ denotes the model parameters.
Correspondingly, the $\chi^2$ function for the $557$ Union2 SNIa
data is given by
 \be\label{eq28}
   \chi^2_\mu({\bf p})=\sum\limits_{i=1}^{557}\frac{\left[\mu_{obs}(z_i)-\mu_{th}(z_i)\right]^2}{\sigma^2(z_i)},
 \ee
where $\sigma$ is the $1\sigma$ error in distance modulus for each
supernova. The parameter $\mu_0$ is a nuisance parameter but it is
independent of the data points. Following \cite{53Nesseris:2005ur},
the minimization with respect to $\mu_0$ is easily performed by
expanding $\chi^2$ of (\ref{eq28}) with respect to $\mu_0$ as
 \be\label{eq29}
   \chi^2_\mu({\bf p})=A-2\mu_0 B+\mu_0^2 C,
 \ee
where
 $$A({\bf p})=\sum\limits_{i=1}^{557}\frac{\left[\mu_{obs}(z_i)-\mu_{th}(z_i;\mu_0=0,{\bf p})\right]^2}{\sigma_{\mu_{obs}}^2(z_i)}\,,$$
 $$B({\bf p})=\sum\limits_{i=1}^{557}\frac{\mu_{obs}(z_i)-\mu_{th}(z_i;\mu_0=0,{\bf p})}{\sigma_{\mu_{obs}}^2(z_i)}, $$
 and
 $$C=\sum\limits_{i=1}^{557}\frac{1}{\sigma_{\mu_{obs}}^2(z_i)}\,.$$
It is evident that (\ref{eq29}) has a minimum for $\mu_0=B/C$ at
 \be\label{eq30}
   \tilde{\chi}^2_{\mu}({\bf p})=A({\bf p})-\frac{B({\bf p})^2}{C}.
 \ee
Since $\chi^2_{\mu,\,min}=\tilde{\chi}^2_{\mu,\,min}$, instead of
minimizing $\chi_\mu^2$ we will minimize $\tilde{\chi}^2_\mu$ which
is independent of the nuisance parameter $\mu_0$. Obviously, the
best-fit value of $h$ can be given by the corresponding $\mu_0=B/C$
of the best fit.

Next, we consider the cosmological observational data from WMAP and
SDSS. For the WMAP data, we use the CMB shift parameter $R$; for the
SDSS data, we use the parameter $A$ of the BAO measurement. It is
widely believed that both $R$ and $A$ are nearly model-independent
and contain essential information of the full WMAP CMB and SDSS BAO
data \cite{55Wang:2006ts}.

The shift parameter $R$ is given by
\cite{54Bond:1997wr,55Wang:2006ts}
 \be\label{eq31}
   R\equiv\Omega_{m0}^{1/2}\int_0^{z_\ast}\frac{d\tilde{z}}{E(\tilde{z})},
 \ee
where the redshift of recombination $z_\ast=1091.3$ which has been
updated in the WMAP 7-year data \cite{56Komatsu:2010fb}. The shift
parameter $R$ relates the angular diameter distance to the last
scattering surface, the comoving size of the sound horizon at
$z_\ast$ and the angular scale of the first acoustic peak in the CMB
power spectrum of temperature fluctuations
\cite{54Bond:1997wr,55Wang:2006ts}. The value of $R$ has been
updated to $1.725\pm0.018$ from the WMAP 7-year data
\cite{56Komatsu:2010fb}. On the other hand, the distance parameter
$A$ from the measurement of the BAO peak in the distribution of SDSS
luminous red galaxies \cite{57Tegmark:2003ud} is given by
 \be\label{eq32}
   A\equiv\Omega_{m0}^{1/2}E(z_b)^{-1/3}\left[\frac{1}{z_b}\int_0^{z_b}\frac{d\tilde{z}}{E(\tilde{z})}\right]^{2/3},
 \ee
where $z_b=0.35$. In \cite{58Eisenstein:2005su}, the value of $A$
has been determined to be $0.469\ (n_s/0.98)^{-0.35}\pm0.017$. Here,
the scalar spectral index $n_s$ is taken to be $0.963$, which has
been updated from the WMAP 7-year data \cite{56Komatsu:2010fb}. So
the total $\chi^2$ is given by
 \be\label{eq33}
   \chi^2=\tilde{\chi}_\mu^2+\chi_{CMB}^2+\chi_{BAO}^2,
 \ee
where $\tilde{\chi}_\mu^2$ is given in (\ref{eq30}),
$\chi_{CMB}^2=(R-R_{obs})^2/\sigma_R^2$ and
$\chi_{BAO}^2=(A-A_{obs})^2/\sigma_A^2$. The best fit model
parameters are determined by minimizing the total $\chi^2$. The
$68.3\%$ confidence level is determined by
$\Delta\chi^2\equiv\chi^2-\chi_{min}^2\leq 1.0$, $2.3$ and $3.53$
for $n_p = 1$, $2$ and $3$, respectively, where $n_p$ is the number
of free model parameters. Similarly, the $95.4\%$ confidence level
is determined by $\Delta\chi^2\equiv\chi^2-\chi_{min}^2\leq 4.0$,
$6.17$ and $8.02$ for $n_p = 1$, $2$ and $3$, respectively.

 \begin{figure*}[htbp]
 \centering \noindent
 \includegraphics[width=7cm]{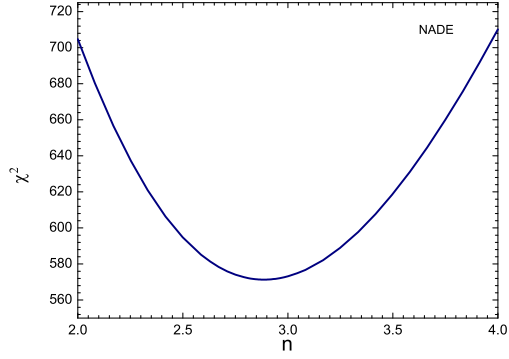}
 \caption{\label{fig1:NADE} (color online) The plot of $\chi^2$ versus $n$ for the NADE model.}
 \end{figure*}



Now, let us discuss the observational constraints on the NADE model.
The NADE model is a single-parameter model whose sole parameter is
$n$. Solving (\ref{eq7}) numerically with the initial condition
$\Omega_q(z_{ini})=n^2(1+z_{ini})^{-2}/4$ at $z_{ini}=2000$ and
substituting the resultant $\Omega_q(z)$ into (\ref{eq4}), the
corresponding $E(z)$ can be obtained.

For the NADE model, the cosmological constraints are found to be:
$n=2.886^{+0.084}_{-0.082}$ at the $1\sigma$ level and
$n=2.886^{+0.169}_{-0.163}$ at the $2\sigma$ level. The best fit
gives values of $\chi^2_{min}=571.338$, $h=0.685$ and
$\Omega_{m0}=0.265$. In Figure~\ref{fig1:NADE}, we plot the relation
of $\chi^2-n$ for the NADE model.

\begin{figure*}[htbp]
 \centering\noindent
 \includegraphics[width=7cm]{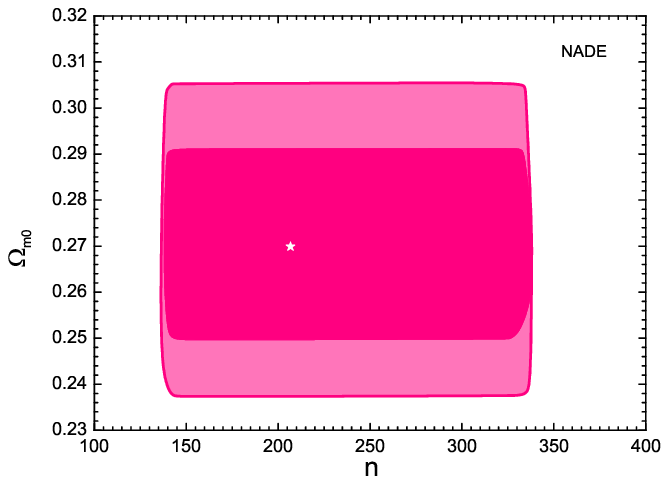}\hskip.5cm
 \includegraphics[width=7cm]{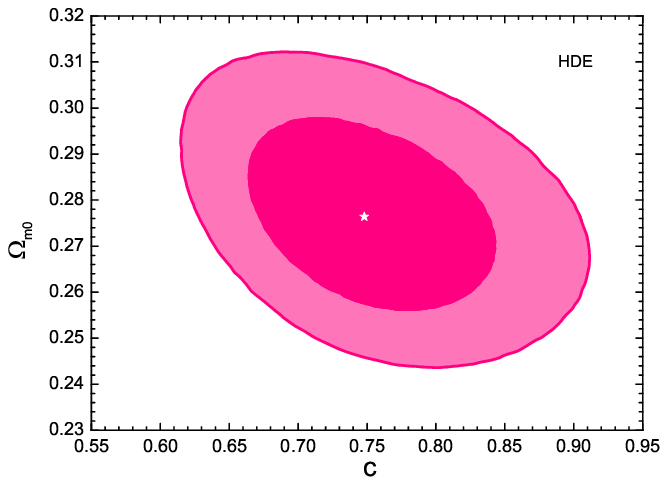}
 \caption{\label{fig3:HDERNADE} (color online) The probability contours at $1\sigma$
 and $2\sigma$ confidence levels in the $\Omega_{m0}-n$ plane for the
 NADE model ({\it Left}) and in the $\Omega_{m0}-c$ plane for the HDE
  model ({\it Right}). Note that here the NADE model is regarded as a two-parameter model. The star denotes the best fit.}
 \end{figure*}

According to \cite{47Wei:2007ty}, the NADE model is always
considered as a single-parameter model, and the initial condition is
chosen to be $\Omega_q(z_{ini})=n^2(1+z_{ini})^{-2}/4$ at
$z_{ini}=2000$. We can, however, adopt the different perspective of
the NADE model being a two-parameter model. In doing so, we are
interested in the result the observational data will tell us. To see
this, we choose the initial condition $\Omega_{q0}=1-\Omega_{m0}$
for the NADE model, and then it becomes a two-parameter model with
the free parameters $\Omega_{m0}$ and $n$. In
Figure~\ref{fig3:HDERNADE} ({\it Left}), we plot contours of the
$68.3\%$ and $95.4\%$ confidence levels in the $\Omega_{m0}-n$ plane
for the NADE model. The fit values for the model parameters are also
shown in Table~\ref{table2}.

\begin{table*} \caption{The fit values for the NADE and HDE models. Note that here the NADE model is regarded as a two-parameter model.}
\begin{center}
\label{table2}
\begin{tabular}{ccccccc}
  \hline\hline
  Model   &                              $\Omega_{m0}$                                      &                                                   $n/c$                           &                           $\chi^2_{min}$        \\
  \hline
  NADE    ~&~~~$0.270^{+0.021}_{-0.020}\left(1\sigma\right)^{+0.036}_{-0.033}\left(2\sigma\right)$~~ & ~~ $206.762^{+131.212}_{-69.060}\left(1\sigma\right)^{+131.212}_{-71.610}\left(2\sigma\right)$     ~~&~~ $542.915$    ~~~  \\
  \hline
  HDE      ~&~~~$0.276^{+0.022}_{-0.020}\left(1\sigma\right)^{+0.036}_{-0.033}\left(2\sigma\right)$~~ & ~~ $0.748^{+0.095}_{-0.085}\left(1\sigma\right)^{+0.164}_{-0.134}\left(2\sigma\right)$            ~~&~~ $543.056$    ~~~  \\
  \hline\hline
\end{tabular}
\end{center}
\end{table*}

 \begin{figure*}[htbp]
 \centering \noindent
 \includegraphics[width=7cm]{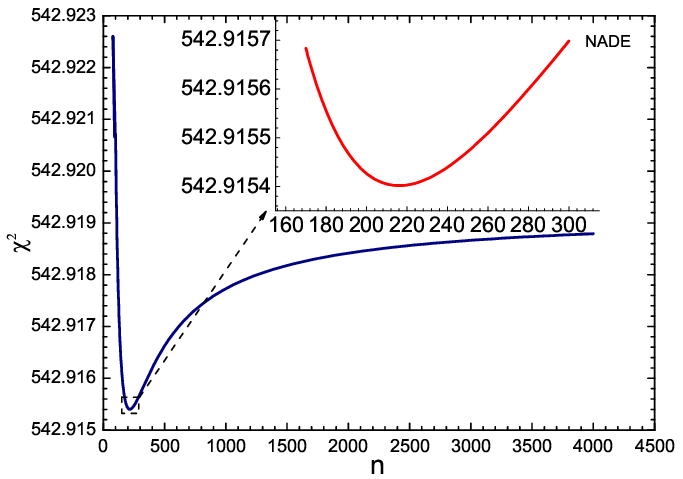}\hskip.5cm
 \includegraphics[width=7cm]{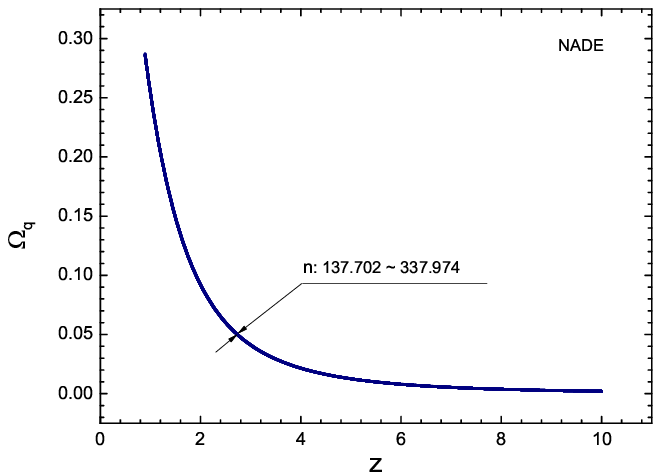}
 \caption{\label{fig4:RNADEtwo} (color online) Left panel: the plot of $\chi^2$ versus $n$ for the two-parameter NADE model
 with $\Omega_{m0}=0.270$. Right panel: the plot of $\Omega_q$ versus $z$ for the two-parameter NADE
 model with $n$ varying from $137.702$ to $337.974$.}
 \end{figure*}

Now, with the NADE model being viewed as a two-parameter model, it
is of great interest to make a direct comparison with the
holographic dark energy (HDE) model. The equation of motion for the
HDE fractional density $\Omega_\Lambda$ is given by
\cite{41Li:2004rb}
 \be\label{eq34}
   \Omega_\Lambda^\prime=\Omega_\Lambda\left(1-\Omega_\Lambda\right)\left(1+\frac{2}{c}\sqrt{\Omega_\Lambda}\right),
 \ee
where $c$ is a numerical parameter similar to $n$ in the NADE model.
Using $\frac{d}{dx}=-\left(1+z\right)\frac{d}{dz}$, we get
 \be\label{eq35}
   \frac{d\Omega_\Lambda}{dz}=-\frac{\Omega_\Lambda}{1+z}\left(1-\Omega_\Lambda\right)\left(1+\frac{2}{c}\sqrt{\Omega_\Lambda}\right).
 \ee
In Figure~\ref{fig3:HDERNADE} ({\it Right}), we plot contours of the
$1\sigma$ and $2\sigma$ confidence levels in the $\Omega_{m0}-c$
plane for the HDE model. The fit values for the model parameters are
also presented in Table~\ref{table2}.

Comparing with the HDE model, we find that the NADE model has a
lower $\chi_{min}^2$. Notwithstanding, it is obviously seen from
Figure~\ref{fig4:RNADEtwo} ({\it Left}) that the data does not
effectively constrain the parameter $n$, i.e., a very large range of
values of $n$ are allowed by the data. Figure~\ref{fig4:RNADEtwo}
({\it Left}) shows the plot of $\chi^2$ versus $n$ with fixed
$\Omega_{m0}$ (which is fixed to the best fit value of 0.270). One
should notice that $\chi^2$ tends to be a constant as $n$ becomes
large and the value of the constant $\chi^2$ is just slightly larger
than $\chi^2_{min}$. It is, therefore, not surprising that $n$ can
range in value from $137.702$ to $337.974$ at the 1$\sigma$ level.
Now, let us discuss the cosmological consequence of such a strong
degeneracy of $n$. For this purpose, we plot the evolution of
$\Omega_q(z)$ with $n$ varying from $137.702$ (the lower limit value
at $1\sigma$) to $337.974$ (the upper limit value at $1\sigma$) in
Figure~\ref{fig4:RNADEtwo} ({\it Right}). We see that the curves
with different $n$ are almost totally degenerate in a narrow region.
This indicates that the cosmological evolution tends to be the same
when $n$ takes large values in this model. In fact, we can also
infer this from (\ref{eq7}). We notice that the term
$\frac{2\left(1+z\right)}{n}\sqrt{\Omega_q}$ is negligible when $n$
is large enough. Thus, (\ref{eq7}) reduces to
 \be\label{eq36}
 \frac{d\Omega_q}{dz}=-\frac{3\Omega_q}{1+z}\left(1-\Omega_q\right).
 \ee
Solving this equation we obtain $\rho_q={\rm constant}$. Therefore,
from the above analysis, we find that when the NADE model is
regarded as a two-parameter model, the dark energy is more likely to
behave as a cosmological constant. So, in the rest of this paper, we
confine our discussions to the single-parameter NADE model.

Next, we discuss the cases of the INADE model. Different cases of
this model are denoted as INADE1, INADE2, INADE3 and INADE4,
respectively. Table~\ref{table1} summarizes the fitting results for
the four cases of the INADE model. For comparison, we also list the
results of the NADE model. In this table we show the best fit,
$1\sigma$ and $2\sigma$ values of the parameters and the
$\chi^2_{min}$ values of the models. The best fit gives values of
$h=0.692$, $0.690$, $0.691$ and $0.688$ and $\Omega_{m0}=0.240$,
$0.236$, $0.237$ and $0.239$ for the four interacting cases. One can
see from Table~\ref{table1} that the INADE1, INADE2 and INADE3
models have a similar $\chi^2_{min}$, and the INADE4 model gives a
larger $\chi^2_{min}$ than the other three, but all are lower than
the NADE model. In addition, a distinctive feature in the INADE4
model is that the fit values of parameter $b$ are all negative in
the $2\sigma$ range. As discussed in the previous section, a
negative $b$ would lead to a negative $\rho_m$ in the future. So, in
this sense, we have shown that the INADE4 model is not a reasonable
model according to the observational data analysis. For the
$1\sigma$ and $2\sigma$ contours in the $n-b$ plane for the four
INADE models, we refer the reader to Figure 5. We will now discuss
the high-$z$ cosmic age problem caused by the old quasar APM
08279+5255 at redshift $z=3.91$.

\begin{table*} \caption{The fit values for the NADE and INADE models. }
\begin{center}
\label{table1}
\begin{tabular}{ccccccc}
  \hline\hline
  Model   &                              $n$                                      &                                                                  $b$                    &                 $\chi^2_{min}$        \\
  \hline
  NADE     ~&~~~$2.886^{+0.084}_{-0.082}\left(1\sigma\right)^{+0.169}_{-0.163}\left(2\sigma\right)$ ~~&~~                       N/A                                                           ~~&~~ 571.338    ~~~  \\
  \hline
  INADE1   ~&~~~$3.199^{+0.194}_{-0.181}\left(1\sigma\right)^{+0.324}_{-0.290}\left(2\sigma\right)$~~ & ~~ $0.029^{+0.008}_{-0.009}\left(1\sigma\right)^{+0.014}_{-0.015}\left(2\sigma\right)$~~&~~ 552.674    ~~~  \\
  \hline
  INADE2   ~&~~~$3.245^{+0.218}_{-0.202}\left(1\sigma\right)^{+0.364}_{-0.322}\left(2\sigma\right)$~~ & ~~ $0.016^{+0.006}_{-0.006}\left(1\sigma\right)^{+0.010}_{-0.010}\left(2\sigma\right)$~~&~~ 556.492    ~~~  \\
  \hline
  INADE3   ~&~~~$3.236^{+0.207}_{-0.193}\left(1\sigma\right)^{+0.344}_{-0.307}\left(2\sigma\right)$~~ & ~~ $0.010^{+0.003}_{-0.004}\left(1\sigma\right)^{+0.006}_{-0.006}\left(2\sigma\right)$~~&~~ 555.015    ~~~  \\
  \hline
  INADE4   ~&~~~$3.208^{+0.239}_{-0.214}\left(1\sigma\right)^{+0.406}_{-0.339}\left(2\sigma\right)$~~ & ~~ $-0.027^{+0.013}_{-0.015}\left(1\sigma\right)^{+0.022}_{-0.025}\left(2\sigma\right)$~~&~~ 561.634   ~~~  \\
  \hline\hline
\end{tabular}
\end{center}
\end{table*}


\section{Age problem: challenge and way out}\label{sec5}

The age of the universe at redshift $z$ is given by
 \be\label{eq37}
   t(z)=\int_z^\infty\frac{dz'}{(1+z')H(z')}.
 \ee
It is convenient to introduce a dimensionless cosmic age
 \be\label{eq38}
   T_{cos}(z)\equiv H_0 t(z)=\int_z^\infty\frac{dz'}{(1+z')E(z')},
 \ee
where $E(z)\equiv H(z)/H_0$. At any given redshift, the age of the
universe should not be less than the age of any object in the
universe, namely, $T_{cos}(z)\geq T_{obj}(z)\equiv H_0 t_{obj}(z)$,
where $t_{obj}(z)$ is the age of an OHRO at redshift $z$. For
convenience, we define a dimensionless quantity, the ratio of the
cosmic and OHRO ages,
 \be\label{eq39}
   \tau(z)\equiv{T_{cos}(z)\over T_{obj}(z)}=H_0^{-1}t_{obj}^{-1}(z)\int_z^\infty \frac{dz'}{(1+z')E(z')}.
 \ee
Thus, the condition $\tau(z)\geq 1$ is equivalent to $T_{cos}(z)\geq
T_{obj}(z)$.

\begin{figure*}[htbp]
 \centering \noindent
 \includegraphics[width=7cm]{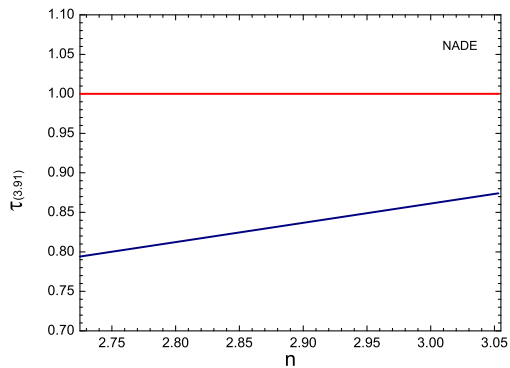}
 \caption{\label{fig5:NADEage} (color online) The plot of $\tau(3.91)$ versus $n$ for the NADE model.
  The region avoiding the age problem is that above the $\tau(3.91)=1$ line (red).} 
 \end{figure*}

First, we will test the NADE model with the ages of the OHROs. We
will use three OHROs: the old galaxy LBDS 53W091 at redshift
$z=1.55$, the old galaxy LBDS 53W069 at $z=1.43$ and the old quasar
APM 08279+5255 at $z=3.91$. The ages of the OHROs at $z=1.55$ and
$z=1.43$ are $3.5$ Gyr and $4.0$ Gyr, respectively. For the age of
the OHRO at $z=3.91$, following \cite{Wei:2007ig}, we use the lower
bound estimate of $t_{obj}(3.91)=2.0$ Gyr. We calculate the age of
the universe at different redshifts using the best fit results of
the NADE model, $n=2.886$ and $h=0.685$, and then we obtain the
values of $\tau$: $\tau(1.55)=1.215$, $\tau(1.43)=1.136$, and
$\tau(3.91)=0.833$, also shown in Table~\ref{table3}. So, the NADE
model can easily accommodate the OHROs at $z=1.55$ and 1.43, but
cannot accommodate the old quasar at $z=3.91$. Of course, the above
result is only for the best fit. We will now see if the old quasar
can be accommodated within the 2$\sigma$ range. We show the result
in Figure~\ref{fig5:NADEage}. In this figure, the blue line
represents $\tau(3.91)$ with $n$ running over the $2\sigma$ range;
for simplicity, we fix $h$ to the best fit value in the calculation.
It is clear that the NADE model does not accommodate the old quasar
APM 08279+5255, therefore, this age problem raises a serious
challenge to the NADE model.

\begin{table*} \caption{The values of $\tau(z)$ in the NADE, INADE1,
INADE2, INADE3 and INADE4 models at best fit concerning LBDS 53W091
at $z=1.55$, LBDS 53W069 at $z=1.43$, and the old quasar APM
08279+5255 at $z=3.91$.}
\begin{center}
\label{table3}
\begin{tabular}{ccccccc}
  \hline\hline
  Model$~\left(n,\ b,\ h\right)$   &   $\tau\left(1.55\right)$  &  $\tau\left(1.43\right)$  &  $\tau\left(3.91\right)$ \\
  \hline
  NADE  ~$\left(2.886,\ 0,\ 0.685\right)$       ~&~~~$1.215$~~ & ~~$1.136$~~ & ~~$0.833$  \\
  \hline
  INADE1~$\left(3.199,\ 0.029,\ 0.692\right)$   ~&~~~$1.326$~~ & ~~$1.238$~~ & ~~$0.919$  \\
  \hline
  INADE2~$\left(3.245,\ 0.016,\ 0.690\right)$   ~&~~~$1.318$~~ & ~~$1.230$~~ & ~~$0.923$  \\
  \hline
  INADE3~$\left(3.236,\ 0.010,\ 0.691\right)$   ~&~~~$1.323$~~ & ~~$1.235$~~ & ~~$0.923$  \\
  \hline
  INADE4~$\left(3.208,\ -0.027,\ 0.688\right)$  ~&~~~$1.290$~~ & ~~$1.204$~~ & ~~$0.908$  \\
  \hline\hline
\end{tabular}
\end{center}
\end{table*}

 \begin{figure*}[htbp]
 \centering \noindent
 \includegraphics[width=16cm]{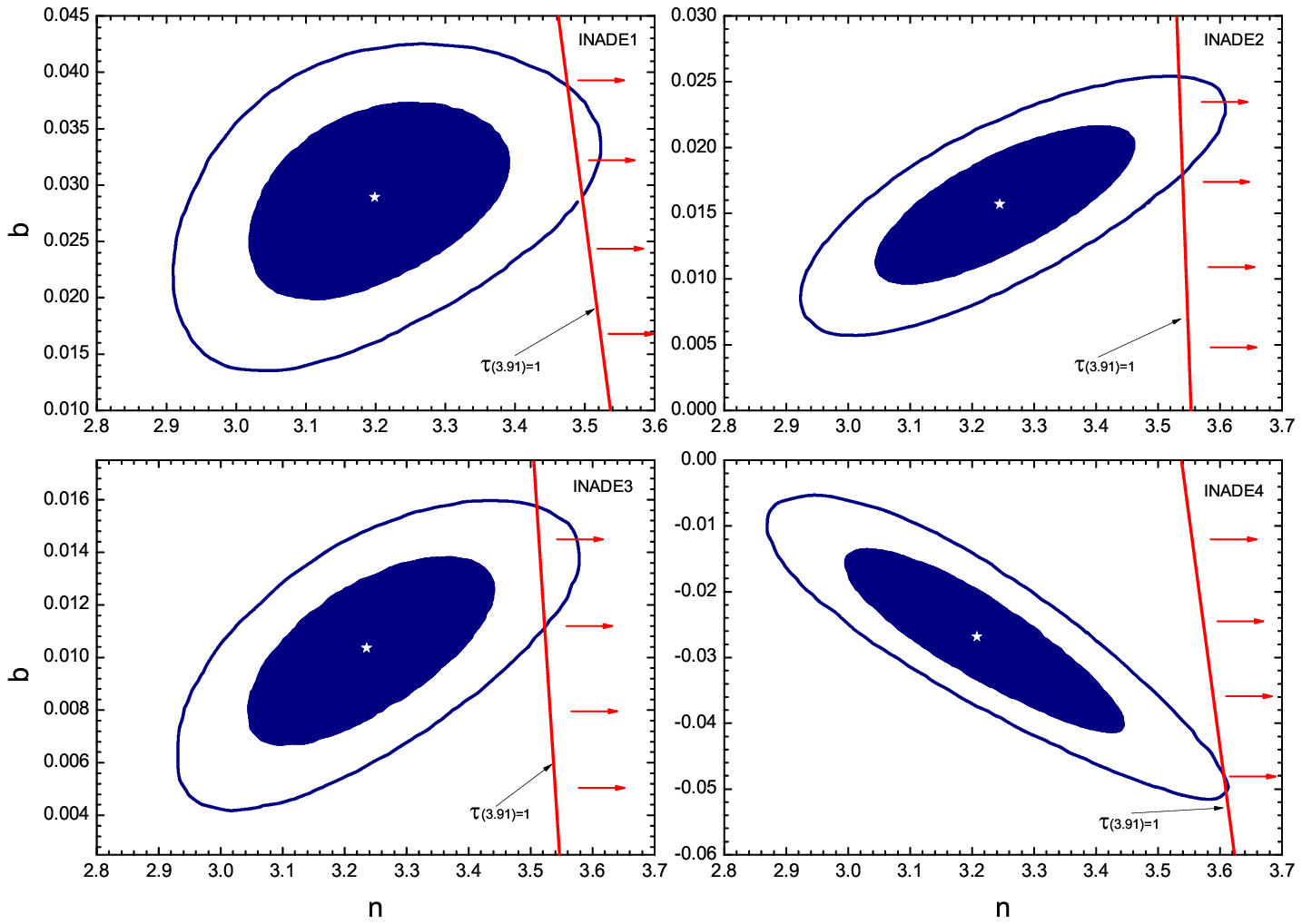}
 \caption{\label{fig6:INADEallage} (color online) The probability contours at the $1\sigma$ and $2\sigma$
 confidence levels in the $b-n$ plane for the four INADE models.
 The red line is an isoline with $\tau(3.91)=1$.
 The allowed region avoiding the age problem is to the right of this line, as indicated by the arrows.}
 \end{figure*}

To overcome this difficulty, we seek help from the possible
interaction between dark energy and dark matter. Thus, we next
explore whether the old quasar APM 08279+5255 can be accommodated in
the INADE model. Based on the above results of observational
constraints, we can easily accomplish this task. Using the best-fit
values, we obtain the values of $\tau(z)$ for the four INADE models,
listed in Table~\ref{table3}. One can see that the $\tau(z)$ values
indeed increase when the interaction is included in the model.
Notwithstanding, for the old quasar at $z=3.91$, the values of
$\tau(3.91)$ only increase from 0.83 to about 0.92, still not
exceeding the value 1. So, if only considering the best fit, the
INADE models cannot solve the age crisis. Of course, it is evident
that the age problem has been greatly alleviated via the interaction
between dark energy and dark matter. As the next step, we will scan
the full parameter space to explore whether there exists an area
able to realize $\tau(3.91)>1$. We show the result in
Figure~\ref{fig6:INADEallage}. The red line indicates where
$\tau(3.91)=1$, using the best fit value for $h$, and thus the
region to the right denotes $\tau(3.91)>1$, as indicated by the
arrows. From this figure, one can clearly see that the red line
passes through the 2$\sigma$ region of the $n-b$ parameter space of
the first three INADE models. For the fourth case, INADE4, the red
line only intersects the edge of the $2\sigma$ region. This result
again shows that the form of interaction $Q=3bH(\rho_q-\rho_m)$ is
not reasonable, since it not only has a negative $b$, but also fails
to provide a solution to the high-$z$ cosmic age problem.

The above analysis indicates that the old quasar APM 08279+5255 can
be successfully accommodated by the INADE model (at least for three
cases) at the $2\sigma$ level. There indeed exists an area within
the 2$\sigma$ scope where $\tau(3.91)>1$ is realized, and thus, it
successfully solves the age crisis of the NADE model. To be modest,
we do not assert that the cosmic age crisis has been completely
solved by the INADE model. After all, only a small area and not the
full region of the 2$\sigma$ scope resides to the right of the
$\tau(3.91)=1$ red line. However, it should be noted that the age
problem has been significantly alleviated by the INADE model under
the current observational constraints.

Finally, we feel that it would be better if some additional comments
on the age problem are given. The age problem is seen in the
apparent old age of a single quasar, APM 08279+5255, whose high Fe/O
ratio derived from X-ray observations has been used to argue for an
age of at least 2 Gyr. At the quasar's redshift of 3.91, the
standard $\Lambda$CDM model is only 1.6 Gyr old and thus apparently
younger than the quasar itself. However, the age estimate of the
quasar rests on the assumption that star formation and evolution
proceeds in a way similar to that observed in the late universe. It
is not even the absolute amount of Fe that serves as an argument,
but the relative Fe/O abundance. Sub-millimeter observations show,
however, that star formation seems to proceed in an atypical way in
APM 08279+5255 \cite{Riechers:2010ft}. This sheds doubt on the age
estimate itself as well as on its reliability. Nevertheless, until a
more accurate measurement of the age of the quasar is available,
theorists have to seriously take the age problem into account.
Likewise, a similar problem of star formation and evolution also
exists for the high-$z$ SNIa, but the observational result of the
cosmic acceleration is still the mainstream. So, it is believed that
the discussion of the age problem in the NADE model is meaningful
and significant at the current stage. Of course, we do expect that
future, more accurate measurements of the age of this old quasar
will naturally eliminate the age crisis in dark energy models.


\section{Conclusion}\label{sec6}

The agegraphic dark energy model originates from the holographic
principle of quantum gravity, so it has physical significance not
only in a phenomenological aspect, but also in a theoretical one.
Though the NADE model has been discussed extensively, it undoubtedly
deserves further investigations. In this paper, we discuss the age
problem in the NADE model.

There is a lot of work addressing the age problem caused by the old
quasar APM 08279+5255 at redshift $z=3.91$, because this quasar has
led many dark energy models into trouble, the $\Lambda$CDM and
holographic dark energy models being no exception. We found in this
paper that the NADE model is also afflicted by the age problem. So,
we explore whether the inclusion of the interaction between dark
energy and dark matter in this model can solve the age problem.

First, we used current observational data to constrain the NADE and
INADE models. For the data, we use the SNIa Union2 sample, the CMB
shift parameter $R$ from 7-yr WMAP, and the BAO parameter $A$ from
the SDSS. We determined the age of the universe in the NADE model by
using the fit results and found that the NADE model cannot realize
$\tau(3.91)>1$ within the full parameter space. We also explored the
cosmological consequence of regarding the NADE model as a
two-parameter model. The fit results show that in such a model the
dark energy is more likely to behave as a cosmological constant.

For the possible interaction between dark energy and dark matter, we
considered four phenomenological cases: $Q=3bH\rho_q$,
$Q=3bH\rho_m$, $Q=3bH(\rho_q+\rho_m)$ and $Q=3bH(\rho_q-\rho_m)$.
The observational data favor a positive $b$ for the three cases, but
give a negative $b$ for the fourth case. We found that when the
interaction is taken into account, the age problem caused by the
quasar can be successfully solved in the first three cases of the
INADE model, at the 2$\sigma$ level. The isoline $\tau(3.91)=1$
passes through the 2$\sigma$ region of the $n-b$ parameter space of
the INADE model. So, there indeed exists an area within the
2$\sigma$ scope realizing $\tau(3.91)>1$ and thus this model
successfully solves the age crisis of the NADE model. Though we
cannot assert that the age crisis has been eliminated in the INADE
model, since most of the parameter space is still in the troubled
area and only the lower bound age of the quasar is used in the test,
we are sure that the age problem has been significantly alleviated
in the INADE model under the current observational constraints.
Therefore, our work can be viewed as further support of the INADE
model.


\begin{acknowledgments}
We would like to thank Xiao-Dong Li for useful discussion and kind
help. This work was supported by the Natural Science Foundation of
China under Grant Nos.~10705041 and 10975032, as well as the
National Ministry of Education of China under the innovation program
for undergraduate students.
\end{acknowledgments}


\end{document}